\documentclass[conference]{IEEEtran}
\usepackage{graphicx}
\usepackage{float}
\usepackage{amsmath,amsfonts,amssymb}
\usepackage{booktabs}
\usepackage{subcaption}
\usepackage[skip=2pt]{caption}
\usepackage[update,prepend]{epstopdf}
\usepackage[lined,algonl,ruled]{algorithm2e}
\usepackage{balance}
\usepackage{acronym}
\usepackage{color}
\usepackage{multirow}
\usepackage{multicol}

\newcommand{\comment}[1]{}

\acrodef{BF}{beamformer}
\acrodef{SPI}{speaker position identifier}
\acrodef{GSC}{general sidelobe canceller}
\acrodef{DSE}{Deep Single Expert}
\acrodef{MVDR}{ minimum variance distortionless response}
\acrodef{GEVD}{generalized eigenvalue decomposition}
\acrodef{AIR}{acoustic impulse response}
\acrodef{PSD}{power spectral density}
\acrodef{cPSD}{cross-power spectral density}
\acrodef{FIR}{finite-impulse response}
\acrodef{MTF}{multiplicative transfer function}
\acrodef{RIR}{room impulse response}
\acrodef{LTI}{linear time invariant}
\acrodef{DNN}{deep neural network}
\acrodef{MFCC}{mel-frequency cepstral coefficients}
\acrodef{MMSE}{minimum mean square error}
\acrodef{ASR}{automatic speech recognition}
\acrodef{ATF}{acoustic transfer function}
\acrodef{LCMV}{linearly constrained minimum variance}
\acrodef{RTFs}{relative transfer functions}
\acrodef{RTF}{relative transfer function}
\acrodef{VAD}{voice activity detector}
\acrodef{STSA}{short-time spectral amplitude estimator}
\acrodef{LSAE}{log-spectral amplitude estimator}
\acrodef{OMLSA}{optimally modified log spectral amplitude}
\acrodef{IMCRA}{improved minima controlled recursive averaging}
\acrodef{STFT}{short-time Fourier transform}
\acrodef{DFT}{discrete Fourier transform}
\acrodef{MoG}{Mixture of Gaussians}
\acrodef{MOE}{Mixture of Experts}
\acrodef{MODE}{Mixture of Deep Experts}
\acrodef{r.v.}{random variable}
\acrodef{p.d.f.}{probability density function}
\acrodef{NN}{neural network}
\acrodef{EM}{expectation-maximization}
\acrodef{SPP}{speech presence probability}
\acrodef{CMVN}{cepstral mean and variance normalization}
\acrodef{NN-MM}{neural network mixture-maximum}
\acrodef{PESQ}{perceptual evaluation of speech quality}
\acrodef{SNR}{signal to noise ratio}
\acrodef{SIR}{signal to interference ratio}
\acrodef{DAE}{deep auto-encoder}
\acrodef{LLR}{log likelihood ratio}
\acrodef{WSS}{weighted spectral slope}
\acrodef{Covl}{overall quality}
\acrodef{Csig}{speech distortion}
\acrodef{Cbak}{background distortion}
\acrodef{WSJ}{wall street journal}
\acrodef{SVM}{support vector machine}
\acrodef{IBM}{ideal binary mask}
\acrodef{IRM}{ideal ratio mask}
\acrodef{ReLU}{rectified linear unit}
\acrodef{WER}{word error rate}
\acrodef{MM}{MixMax}
\acrodef{MOS}{mean opinion score}
\acrodef{mse}{mean squre error}
\acrodef{pDNN}{phoneme DNN}
\acrodef{cDNN}{classifier DNN}
\acrodef{SGD}{stochastic gradient descent}
\acrodef{TDOA}{time difference of arrival}
\acrodef{CSD}{concurrent speakers detector}
\acrodef{STOI}{short-time objective intelligibility measure}
\acrodef{MCCSD}{multichannel CSD}

\newcommand*\justify{%
  \fontdimen2\font=0.4em
  \fontdimen3\font=0.2em
  \fontdimen4\font=0.1em
  \fontdimen7\font=0.1em
  \hyphenchar\font=`\-
}

\usepackage{amsmath,amssymb,amsfonts}
\usepackage{algorithmic}
\usepackage{graphicx}
\usepackage{booktabs}
\usepackage{subcaption}
\usepackage{textcomp}
\usepackage{caption}

\begin{document}

\title{Speech Dereverberation Using Fully Convolutional Networks}

\author{\IEEEauthorblockN{Ori Ernst, Shlomo E. Chazan, Sharon Gannot and Jacob Goldberger}
\IEEEauthorblockA{Faculty of Engineering
Bar-Ilan University\\
Ramat-Gan, 5290002, Israel.\\
{\{Ori.Ernst, Shlomi.Chazan, Sharon.Gannot, Jacob.Goldberger\}@biu.ac.il }}
}

\maketitle

\begin{abstract}
Speech derverberation using a single microphone is addressed in this paper. 
Motivated by the recent success of the fully convolutional networks (FCN) in many image processing applications, we investigate their applicability to enhance the speech signal represented by short-time Fourier transform (STFT) images. We present two variations: a ``U-Net" which is an encoder-decoder network with skip connections and a generative adversarial network (GAN) with U-Net as generator, which yields a more intuitive cost function for training. To evaluate our method we used the data from the REVERB challenge, and compared our results to other methods under the same conditions. We have found that our method outperforms the competing methods in most cases.

\end{abstract}


\section{Introduction}
Reverberation, resulting in from multiple reflections from the rooms facets and objects, degrade the speech quality, and in severe cases, the speech intelligibility, especially for hearing impaired people. The success rate of automatic speech recognition (ASR) systems may also significantly deteriorate in reverberant conditions, especially in cases of mismatch between the training and test phases.
Reverberation is the result of convolving an anechoic speech utterance by a long acoustic path. The output signal suffers from overlap- and self-masking effects that may deteriorate the speech quality~\cite{nabvelek1989reverberant}. These are often manifested as ``blurring'' effects on the short-time Fourier transform (STFT) images.
A plethora of methods for speech dereverberation using both single- and multi-microphone exists~\cite{naylor2010speech}.

The REVERB challenge~\cite{reverberation_challenge} was a large community-wide endeavor to put together a common dataset for testing dereverberation algorithms and for evaluating many algorithms on a common ground.
The challenge addresses both ASR and speech enhancement using 1, 2 or 8 microphones. In this paper, we focus on single-microphone speech enhancement solutions.

In STFT domain, reverberation can be modeled as a per frequency convolution along frames. In~\cite{WPE}, the signal is dereverberated by estimating the inverse filter that minimizes the weighted linear prediction error (WPE). A different approach directly estimates the reverberant filter, using recursive expectation-maximization algorithm (REM), that is then used to construct a Kalman filter. This method can be both applied in the single- and multi-microphone cases~\cite{KEM}.

Other methods in the challenge used a convolution in the amplitude of the STFT domain and then applied nonnegative
matrix factor deconvolution (NMFD)~\cite{NMF2}.

Spectral domain processing can be also utilized by using the statistical model of the reverberant tail (see Polack's model~\cite{polack1993playing}). A method that takes into account the direct-to-reverberant ratio (DRR) was presented in~\cite{habets2009late}. A modified version of this approach, proposed by Cauchi et al.~\cite{Cauchi} proved to be very efficient in the single-microphone case.
Two additional methods used spectral enhancement procedure, using spectral analysis tools other than the STFT. Gonzalez et al.~\cite{Gonzalez} used a zero-phase transformation, which can distinguish between a periodic and a non-periodic components of the speech signal. Wisdom et al.~\cite{Wisdom} proposed a short-time fan-chirp transform (STFChT) that is coherent with speech signals over a longer analysis window.
Deep learning methods were employed by Xiao et al.~\cite{Xiao} to find a nonlinear mapping between the reverberant and clean spectrum.

Following the successes of learning-based methods, Han et al.~\cite{wang15} applied a DNN to map noisy and reverberant spectrograms to clean spectrograms. Then, at a post-processing stage, an iterative phase reconstruction was employed to reconstruct the time-domain signal. Williamson et al.~\cite{wang17} used  deep learning techniques to directly estimate the complex ideal ratio mask (cIRM).
Weninger et al.~\cite{LSTM} used bi-directional long short-term memory (bi-LSTM) recurrent neural networks (RNNs) to preserve the speech continuity.

Despite the success of fully connected (FC) and LSTM networks, they do not fully utilize the spectral structure of the speech. In the speech spectrogram, there are clear time-frequency patterns that can be exploited. By dividing the spectrum into time frames, these patterns are not preserved.

Another type of network is known as the convolutional neural network (CNN), which is based on a sliding-window process in order to enhance the current time frame. In CNN, each pixel in the target image is computed using only a small number of context pixels from the original image, followed by a FC layer that ignores any existing time-frequency structure~\cite{CNN}.

In scenery images, CNN is usually used for classification tasks. For segmentation tasks, which necessitates an estimate the entire picture, a  fully convolutional network (FCN)~\cite{FCN} is most commonly used.

In this study, we apply FCN architecture to the speech dereverberation task. We show that this approach, which preserves global temporal and spectral information along with local information, significantly outperforms competing methods, as demonstrated on both real data and (most of) the simulated data in REVERB challenge.


\section{Problem formulation}

Reverberant speech can be modeled as
\begin{equation}
z(t) = \{x*h\}(t)\label{reverb_model}
\end{equation}
where $x$ is the clean speech that is convolved by $h$, the room impulse response (RIR) that creates the reverberation. In REVERB challenge, a low-level stationary noise is added to $z(t)$, but it is neglected in our derivations. We would like to retrieve $x(t)$ from $z(t)$ by a nonlinear function that is implemented by a neural network:
\begin{equation}
x(t) \approx f(z(t))\label{non-linear_func}.
\end{equation}
Following~\cite{log_spec} we use the log-spectrum as an effective feature vector. Let $Z(n,k)$ denote the log-absolute value of the STFT of $z(t)$ in the $n$-th time-frame and the $k$-th frequency bin. Let $L$ denote the frame-length of the transform, hence, due to symmetry, the indexes of the log-spectrum are $k=0,...,L/2$. Similarly, $X(n,k)$ denote the log-spectrum of the clean speech.
Fig.~\ref{subfig:noisy} and Fig.~\ref{subfig:clean} depict an example of a reverberant and a clean log-spectrogram, respectively. Comparing the two figures, the reverberant spectrograms is much more ``smeared" than the clean spectrogram, as a direct consequence of the convolution with the long RIR.

%

\section{A U-Net based network architectures}
Following the speech enhancement method in~\cite{noise_gan}, the time-frequency (T-F) representation, the spectrogram, can be treated as an image. Consequentially, the enhancement task becomes an image-to-image transformation.
Treating the reverberant speech as an image has two major advantages. First, speech spectrograms exhibit typical
patterns (e.g. pitch continuity, harmonic structure, and formants). An image processing methodology can take advantage of these structures to apply relevant enhancement procedure. Second, this representation allows us to use highly successful image transformation methods such as the fully convolutional network (FCN).

In this section, we present three variants of the proposed algorithm, all based on a U-Net architecture, namely the U-Net image-2-image architecture with two filter shapes, and U-net on conjunction with generative adversarial network (GAN).

\subsection{U-Net Image-2-Image Architecture}
\label{sec:Unet}

In this study we propose an FCN as the basic architecture. 
In this network the image is downsampled and upsampled again, which causes a rapid increase in the receptive field that serves to propagate global information in both time and frequency axes. The receptive field of a neuron is the number of pixels from input image that are used for computing the neuron value. 
In FCN the image is downsampled until a bottleneck of $1 \times 1$, causing each pixel in the target to be influenced by the entire input image. 

\begin{figure*}[htbp]
\centering
\begin{subfigure}[b]{0.3\linewidth}
	\centering
   \includegraphics[width=.90\linewidth]{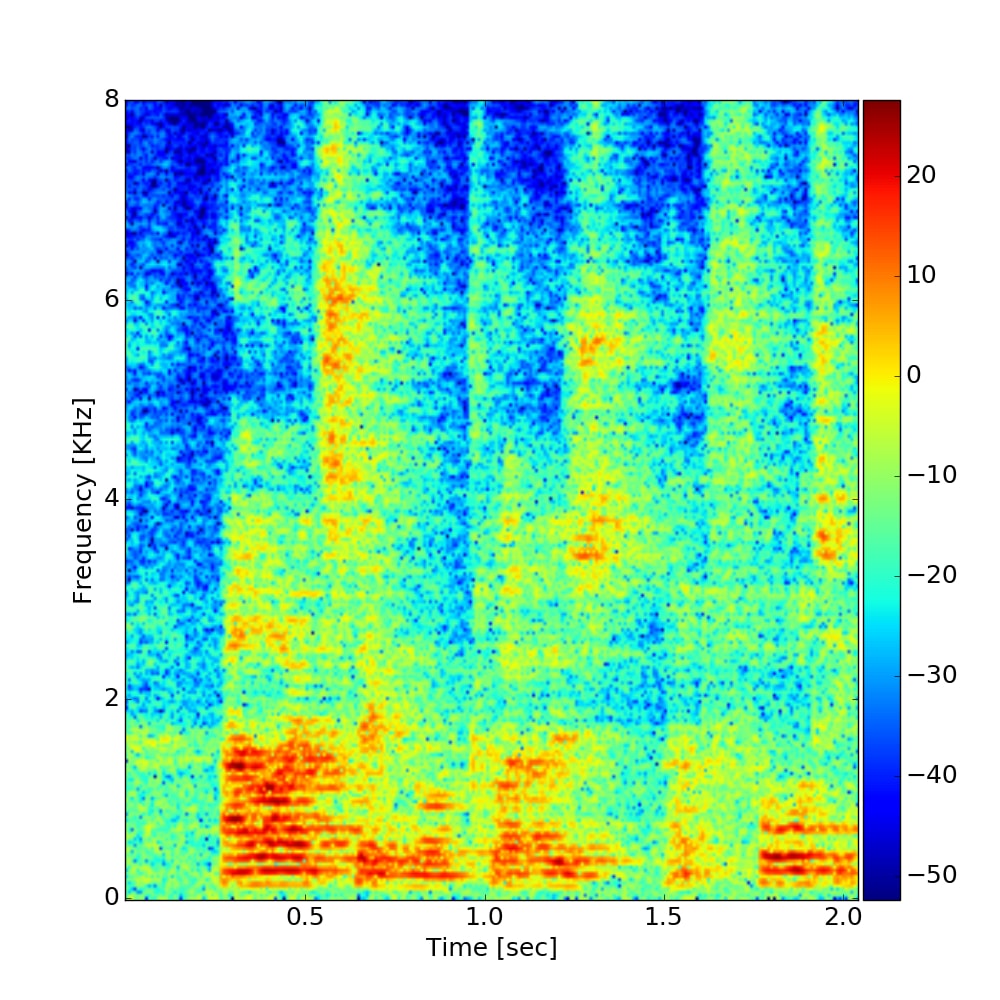}
  \caption{Noisy and Reverberant log STFT}
  \label{subfig:noisy}
  \end{subfigure}
\begin{subfigure}[b]{0.3\linewidth}
	\centering
    \includegraphics[width=.90\linewidth]{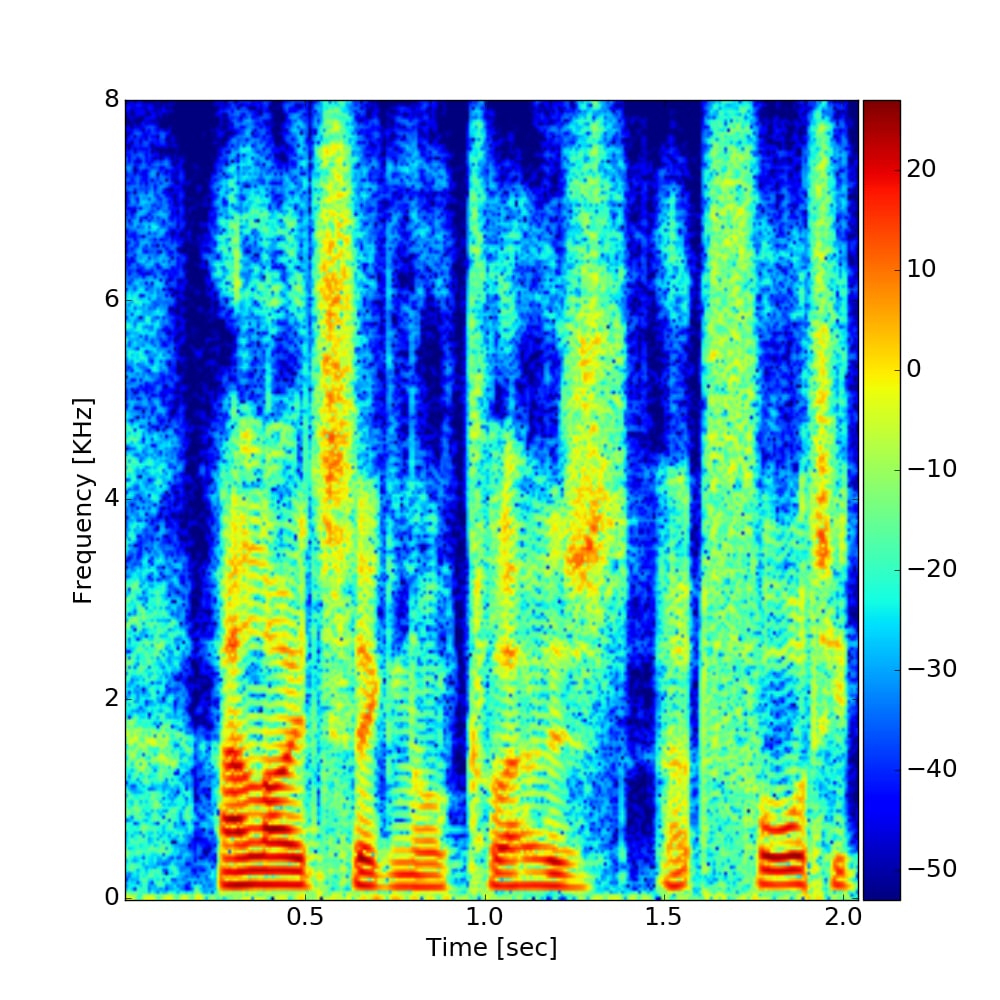}
  \caption{Clean log STFT}
  \label{subfig:clean}
 \end{subfigure}
\begin{subfigure}[b]{0.3\linewidth}
	\centering
   \includegraphics[width=0.90\linewidth]{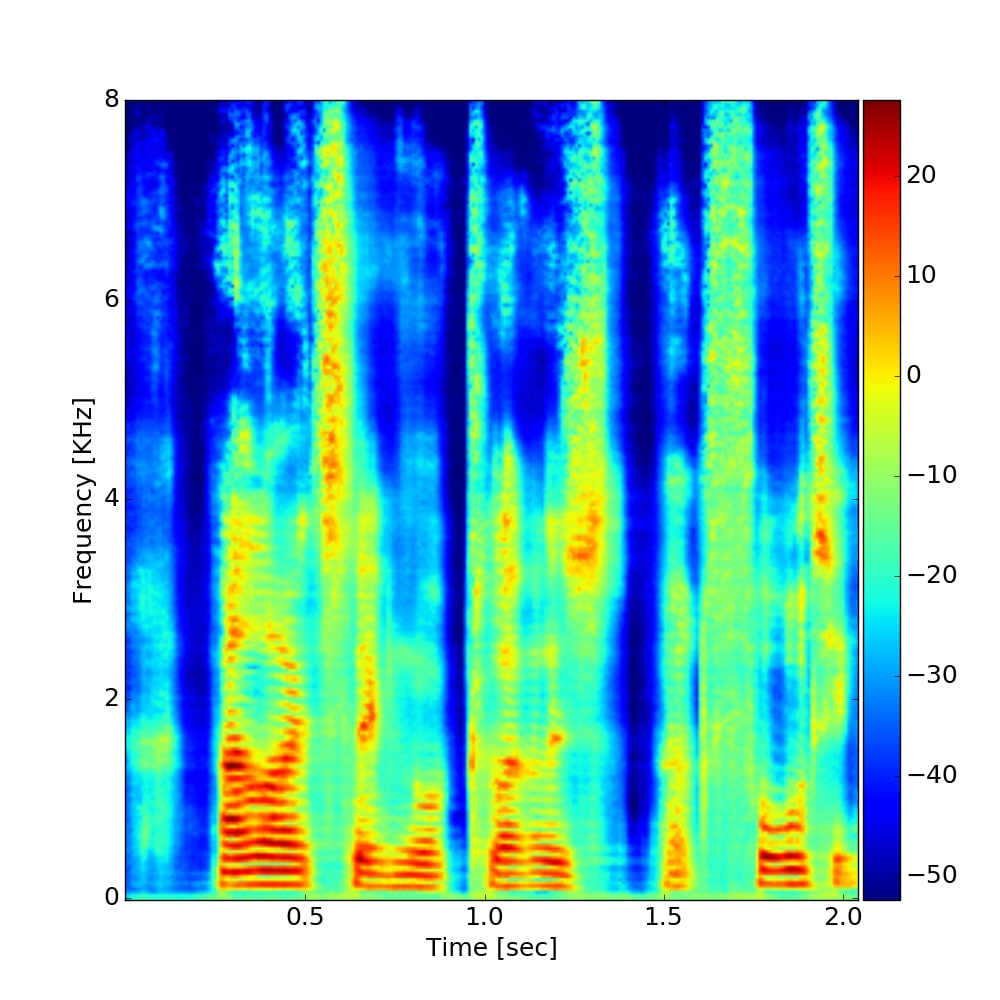}
  \caption{U-Net with asymmetric filters result}
  \label{subfig:unet_unsym}
\end{subfigure}

\centering
\begin{subfigure}[b]{0.3\linewidth}
	\centering
   \includegraphics[width=.90\linewidth]{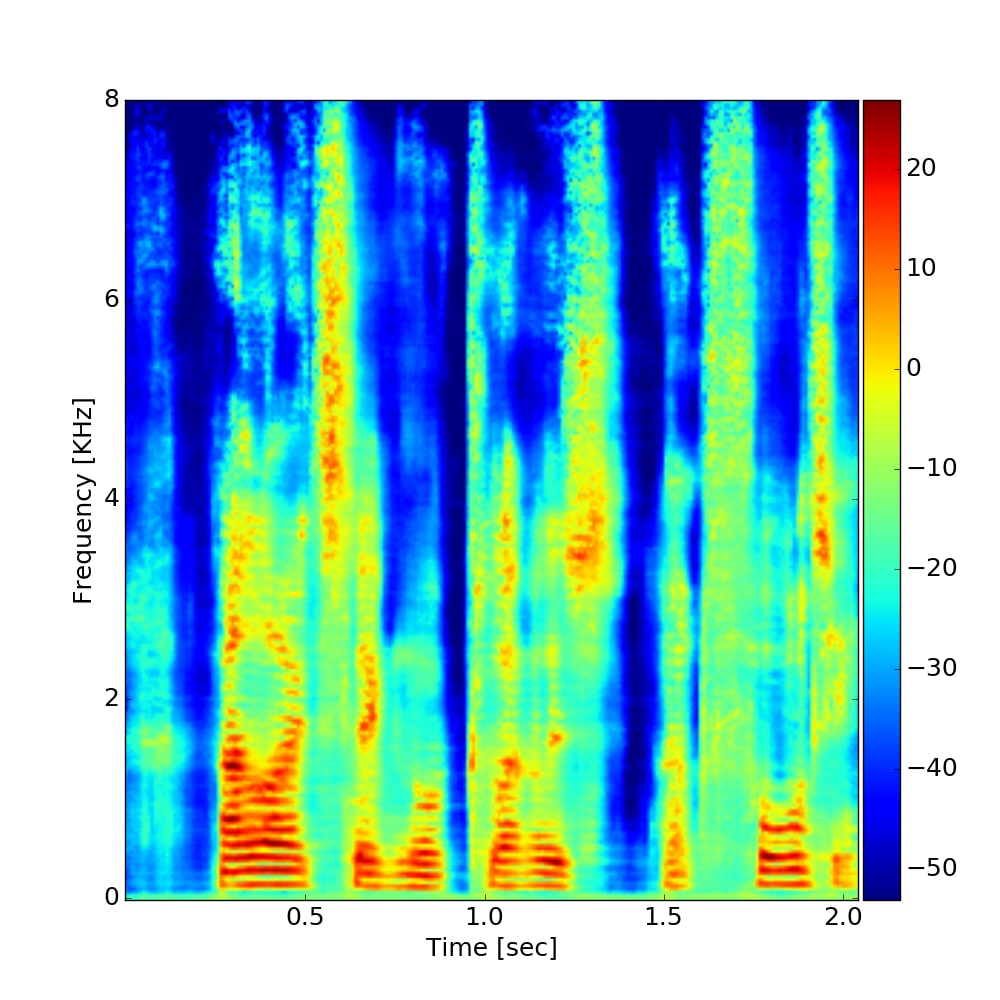}
  \caption{U-Net with symmetric filters result}
  \label{subfig:unet_sym}
\end{subfigure}
\begin{subfigure}[b]{0.3\linewidth}
	\centering
   \includegraphics[width=.90\linewidth]{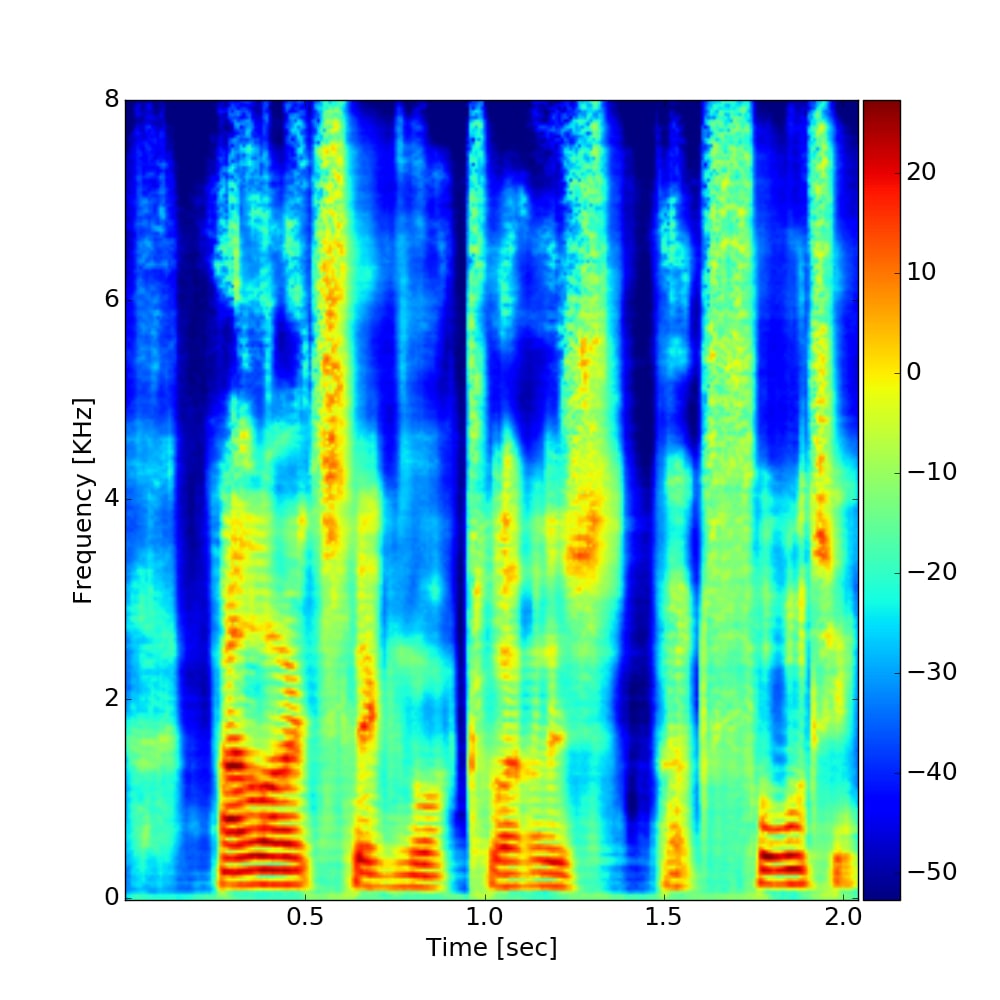}
  \caption{U-Net + GAN result}
  \label{subfig:unet_gan}
\end{subfigure}
\caption{One example of reverberant and clean log-spectrograms, and results applied by all methods.}
\label{fig:example}
\end{figure*}

An encoder-decoder network is a very common method for image-to-image translation. In this type of network, each layer downsamples its input (usually with a stride of 2) to the next layer until there is a bottleneck. In the subsequent layer, the input goes through the reverse process where each layer upsamples its input until it returns to the original shape. Thus, the network input is a high resolution image that is squeezed to a very low resolution image (a bottleneck). Conversely, the expanding path does the opposite, i.e., it increases the image resolution until it is resized to the original dimensions. Unfortunately, this network tends to loose essential low level information during the downsampling procedure. In order to improve the encoder-decoder structure, the ``U-Net" architecture \cite{u-net}, along with its symmetric squeezing and expanding paths (like a ``U" shape), capitalizes on the fact that input and output images should have the same structure. We can thus circumvent the bottleneck and also transmit the shared information without downsampling. Therefore, the U-net connects between mirrored layers in the encoder and decoder stacks that transmit the information without going through a bottleneck. In other words, the skip connections directly concatenate feature maps from layer $i$ in the encoder to layer $N-i$ in the decoder, where $N$ is number of layers.

Following \cite{noise_gan}, our network details are as follows.
Let $\textrm{CBL}_{l,s}$ denote a Convolution-BatchNorm-Leaky-ReLU layer with slope=0.2, where $l$ is number of filters and $s \times s$ is the filter size. $\textrm{CL}_{l,s}$ and $\textrm{CBR}_{l,s}$ have the same architecture but without BatchNorm, or with a non-leaky ReLU, respectively. With same notation, let $\textrm{DCDR}_{l,s}$ denote the DeConvolution-BatchNorm-Dropout-ReLU with dropout of 50\%, and let $\textrm{DCR}_{l,s}$ denote the DeConvolution-BatchNorm-ReLU. $\textrm{DCT}_{l,s}$ denote DeConvolution-tanh.
The U-Net squeezing path is given by:\\
$\textrm{CL}_{64,5} \rightarrow \textrm{CBL}_{128,5} \rightarrow \textrm{CBL}_{256,5} \rightarrow \textrm{CBL}_{512,5} \rightarrow \textrm{CBL}_{512,5} \rightarrow \textrm{CBL}_{512,5} \rightarrow \textrm{CBL}_{512,5} \rightarrow \textrm{CBR}_{512,5}$\\
and the U-Net expanding path is:\\
$\textrm{DCDR}_{512,5} \rightarrow \textrm{DCDR}_{512,5} \rightarrow \textrm{DCDR}_{512,5} \rightarrow \textrm{DCR}_{512,5} \rightarrow \textrm{DCR}_{256,5} \rightarrow \textrm{DCR}_{128,5} \rightarrow \textrm{DCR}_{64,5} \rightarrow \textrm{DCT}_{1,5}$.\\

\begin{figure}[htbp]
\centerline{\includegraphics[width=\linewidth]{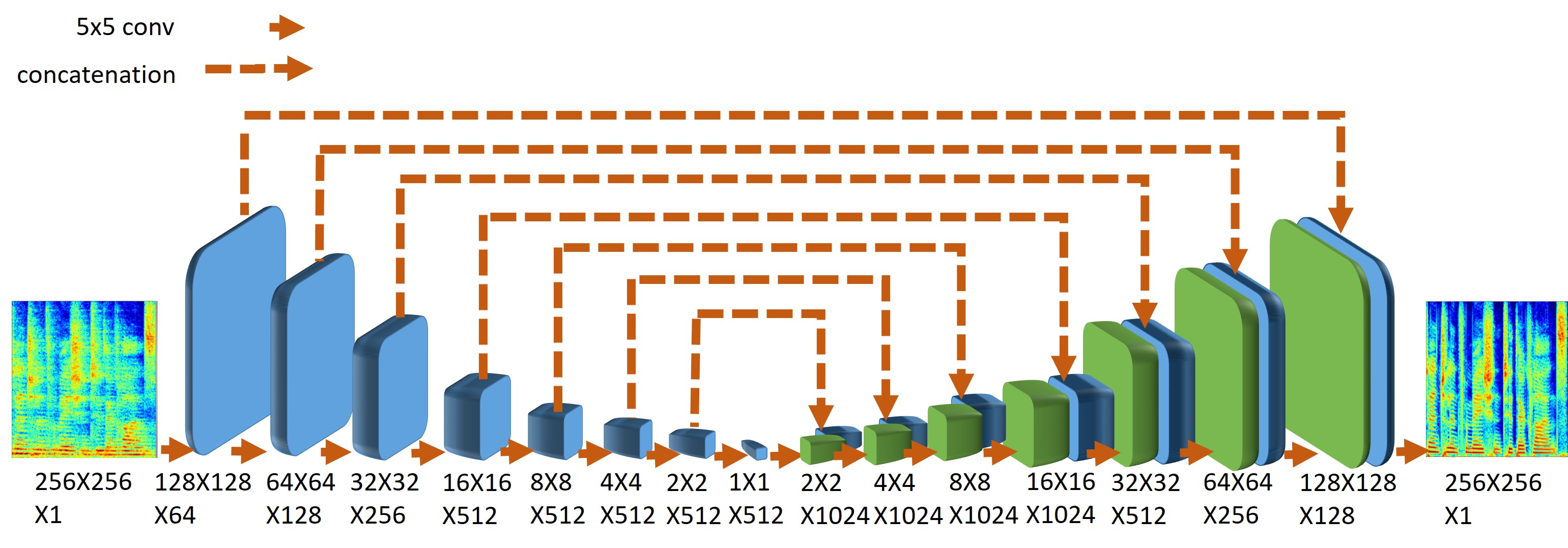}}
\caption{U-Net for speech dereverberation architecture.}
\label{fig:unet}
\end{figure}

The U-net architecture which combines direct and skip connections is illustrated in Fig. \ref {fig:unet}. The input is normalized to the range [{-}1,1] before processing.
At the end of the network, tanh is applied in order to confine the output $\hat{X}(n,k)$ to the range [{-}1,1], same as the input.

U-Net filters are usually symmetric ($5 \times 5$ in our implementation), but this is not necessarily the optimal choice. In regular images, using symmetric filters makes sense because there is no difference between the $x$ and $y$ axes. Nevertheless, our images are actually spectrograms, with one axis representing the time domain, and the other the frequency domain. To mitigate the reverberation effects, it may be better to use higher-dimension in the frequency-domain than in the time-domain, to capture important spectral patterns of the speech. The pitch structure occupies several frequency bands, so a filter must be at least longer than the fundamental pitch frequency. Accordingly, we investigated the use of asymmetric filters of size  $10 \times 5$ pixels, 10 for the frequency domain, and 5 for the time domain. In most cases, as we demonstrate in the experimental section, these filters exhibit better performance measures than the symmetric filters. Examples for the log-spectrograms of the enhanced speech by U-Net and asymmetric U-Net are shown in Fig.~\ref{subfig:unet_sym} and Fig.~\ref{subfig:unet_unsym}, respectively.


To train the network we used a training dataset that consisted of T-F images of the clean speech $X_t(n,k)$ and the corresponding noisy and reverberant signal $Z_t(n,k)$, to generate the enhanced speech $\hat{X}_t(n,k)$, whereas $t=0,...,T-1$ represents the $t$-th example and $T$ is the number of training data examples.
Following~\cite{noise_gan} and~\cite{u-net}, we trained the U-Net with the Adam optimizer for 10 epochs, with a batch size of a single image. The cost we minimize is the mean square error (MSE). The loss function we minimize at the training is thus:
\begin{equation}
L_{\textrm{MSE}} = \sum_{t} \|X_t-\hat{X}_t\|^2 \label{L2}
\end{equation}
where $t$ goes over all the images of the training data. Although in~\cite{pix2pix} the loss function used was $L_1$ norm, here $L_2$ was found to yield better results.



\begin{table*}[h!]
\centering
\setlength{\belowcaptionskip}{.1cm}

\caption{Results of simulated data for far microphones.}
\label{table:sim_far}
\begin{tabular}{|l|ccc|ccc|ccc|ccc|}
\hline
 & \multicolumn{3}{c|}{CD} & \multicolumn{3}{c|}{LLR} & \multicolumn{3}{c|}{FWSegSNR} & \multicolumn{3}{c|}{SRMR} \\  Room
 &  1 &  2 & { 3} &  1 &  2 & { 3} & 1 &  2 & 3 &  1 &  2 &  3 \\   \hline
{reverberant speech} & 2.67 & 5.21 & {4.96} & 0.38 & 0.75 & {0.84} & 6.68 & 1.04 & {0.24} & 4.58 & 2.97 & 2.73 \\
{Cauchi et al. \cite{Cauchi}} & 2.67 & 4.65 & {4.44} & 0.42 & 0.77 & {0.82} & 8.93 & 3.50 & {2.75} & 4.75 & 3.88 & 3.86 \\
{Gonzalez et al. \cite{Gonzalez}} & 3.59 & 5.03 & {5.15} & 0.31 & \textbf{0.54} & {0.65} & 5.72 & 2.74 & {1.64} & \textbf{5.98} & 4.20 & 3.86 \\
{Wisdom et al. \cite{Wisdom}} & 2.83 & 4.53 &{4.49} & 0.37 & 0.79 &{0.79} & 8.86 & 5.01 & 3.75 & 5.30 & 3.93 & 3.63 \\
{Xiao et al. \cite{Xiao}} & \textbf{1.92} & \textbf{3.17} & {2.99} & 0.41 & 0.61 & {0.58} & 9.12 & 6.31 & {5.97} & 5.67 & \textbf{5.80} & \textbf{5.03} \\
{U-Net} & 2.06 & 3.41 & {3.05} & \textbf{0.26} & 0.63 & {0.58} & 11.80 & 8.60 & {8.65} & 4.98 & 5.44 & 4.79 \\
{asymmetric U-Net} & 2.09 & 3.24 & {2.96} & \textbf{0.26} & 0.57 & {\textbf{0.55}} & 11.96 & 8.90 & {9.02} & 4.83 & 5.32 & 4.65 \\
{asymmetric U-Net + GAN} & 2.05 & 3.19 & {\textbf{2.92}} & \textbf{0.26} & 0.57 & {0.56} & \textbf{12.08} & \textbf{9.00} & {\textbf{9.05}} & 4.76 & 5.27 & 4.71 \\ \hline
\end{tabular}%
\end{table*}

\begin{table*}[h!]
\centering
\caption{Results of simulated data for near microphones.}
\label{table:sim_near}
\begin{tabular}{|l|ccc|ccc|ccc|ccc|}
\hline
 & \multicolumn{3}{c|}{CD} & \multicolumn{3}{c|}{LLR} & \multicolumn{3}{c|}{FWSegSNR} & \multicolumn{3}{c|}{SRMR} \\   Room
 &  1 &  2 &  3 &  1 &  2 &  3 &  1 &  2 &  3 &  1 &  2 &  3 \\  \hline
{reverberant speech} & 1.99 & 4.63 & {4.38} & 0.35 & 0.49 & {0.65} & 8.12 & 3.35 & {2.27} & 4.50 & 3.74 & 3.57 \\
{Cauchi et al. \cite{Cauchi}} & 2.02 & 3.82 & {3.67} & 0.36 & 0.51 & {0.64} & 10.29 & 6.19 & {4.89} & 4.65 & 4.32 & 4.27 \\
{Gonzalez et al. \cite{Gonzalez}} & 3.24 & 4.53 & {4.76} & 0.26 & \textbf{0.34} & {0.50} & 7.13 & 5.13 & {3.96} & \textbf{6.05} & 5.45 & 5.01 \\
{Wisdom et al. \cite{Wisdom}} & 2.29 & 3.64 & {3.65} & 0.31 & 0.54 & {0.60} & 10.07 & 8.24 & {6.51} & 5.18 & 4.70 & 4.56 \\
{Xiao et al. \cite{Xiao}} & \textbf{1.58} & 2.65 & {2.68} & 0.37 & 0.50 & {0.52} & 9.79 & 7.27 & {6.83} & 5.74 & \textbf{6.49} & \textbf{5.86} \\
{U-Net} & 1.73 & 2.71 & {2.61} & \textbf{0.19} & 0.44 & {\textbf{0.45}} & \textbf{13.33} & 10.71 & {10.18} & 4.73 & 5.35 & 5.05 \\
{asymmetric U-Net} & 1.79 & 2.63 & {2.54} & 0.20 & 0.42 & {\textbf{0.45}} & 13.32 & 10.83 & {\textbf{10.45}} & 4.48 & 5.07 & 4.90 \\
{asymmetric U-Net + GAN} & 1.75 & \textbf{2.58} & {\textbf{2.53}} & 0.20 & 0.41 & {\textbf{0.45}} & 13.32 & \textbf{10.87} & {10.40} & 4.51 & 5.09 & 4.94 \\ \hline
\end{tabular}%
\end{table*}

\begin{table}[t]
\centering
\caption{SRMR Results of real data for far and near microphones.}
\label{table:real}
\begin{tabular}{|l|c|c|}
\hline                                                     method            &  far    &  near   \\ \hline
{reverberant speech}  & 3.19          & 3.17          \\
{Cauchi et al. \cite{Cauchi}}       & 4.76          & 4.87          \\
{Gonzalez et al. \cite{Gonzalez}}     & 4.62          & 4.78          \\
{Wisdom et al. \cite{Wisdom}}       & 4.82          & 4.96          \\
{Xiao et al. \cite{Xiao}}         & 4.42          & 4.29          \\
{U-Net}               & 5.54          & 5.45          \\
{asymmetric U-Net} & \textbf{5.68}   & \textbf{5.47} \\
{asymmetric U-Net + GAN}              & 5.52          & 5.34          \\ \hline
\end{tabular}
\end{table}

\subsection{GAN Training Strategy}
Minimizing the MSE between the enhanced and the clean speech is not always aligned with the human judgment. For that reason, generative adversarial networks (GAN)~\cite{GAN} have been successfully applied to several image processing tasks. The pix2pix conditional GAN (cGAN)~\cite{pix2pix} that presents a way to perform image translations (such as B\&W to color images) using GAN, was found specifically attractive. This method was used by~\cite{noise_gan} for a noisy speech enhancement task, but it has never been applied to a reverberation task. Our cGAN was composed of two components: a \emph{generator} (G) that enhances the spectrogram (the network described in Sec.~\ref{sec:Unet}), and a \emph{discriminator} (D) that was trained to distinguish between the result of G and a clean spectrogram. The discriminator receives two images. The first is the output of G or a clean image, and the second is the noisy spectrogram used as a condition. During training, the goal of G was to improve itself, such that D would not be able to distinguish between the output of G  and the clean spectrogram. The objective is the same as in~\cite{noise_gan} and~\cite{pix2pix}:
\begin{equation}
\begin{aligned}
L_{\textrm{GAN}} (G,D) &=  \sum_t (\log D(Z_t,X_t) + \log (1- D(Z_t,G(Z_t)))\label{gan_loss}
\end{aligned}
\end{equation}
such that $Z_t$, $X_t$ and $\hat{X}_t=G(Z_t)$ are the $t$-th example of the reverberant, clean and enhanced log-spectrum images respectively.
To improve the results, the MSE loss was added to the GAN loss as a regularization term that ensures that the enhanced speech is close to the clean speech.
Accordingly, the final GAN score was expressed as
\begin{equation}
L (G,D) = L_{\textrm{GAN}} (G,D) +\lambda L_{\textrm{MSE} } (G)\label{full_loss}
\end{equation}
where $\lambda$ is the weight of the direct MSE loss.
The GAN network was initialized with the U-Net weights, and then was trained for a couple of more epochs.
We have found empirically the $\lambda=1000$ yields good results. An example of GAN based dereverberation is depicted in Fig.~\ref{subfig:unet_gan}. Both U-Net and GAN architectures are based on~\cite{noise_gan} implementation.

\section{Experimental results}

We implemented the three methods described above: U-Net, asymmetric U-Net, and GAN initialized with asymmetric U-Net. We calculated all measurements on the test dataset by the REVERB challenge published script, so the comparison to other methods would be fair.
\subsection{Dataset}
Our data were based on the REVERB challenge \cite{reverberation_challenge}. The data were divided into a training set and an evaluation test set, such that the former only included simulated data, whereas the latter also included real recordings. The simulated data were taken from the WSJCAM0 corpus \cite{WSJCAMO}, in which each utterance was convolved randomly with room impulse responses (RIRs) from different rooms and a noise was added at an SNR of 20dB. The simulated test dataset was generated from three different room sizes (small, medium and large) with a reverberation time ($T_{60}$) of approximately 0.25s, 0.5s and 0.7s, respectively, and from two microphone placements relative to the speaker (200cm and 50cm). The real recordings were taken from the MC-WSJ-AV corpus \cite{MC-WSJ-AV}, which contains recordings from a noisy and reverberant meeting room, with $T_{60}$ of approximately 0.7s, and microphone distance of 250cm and 100cm from the speaker. The training data were generated from 24 ``rooms" that were simulated from suitable RIRs with reverberation times changed from 0.2s to 0.8s. The rooms and conditions were different for the evaluation set and the training set.

\subsection{Pre-Processing} The input to the network was the log-spectral image of the noisy and reverberant speech. STFT was computed using frame length of 512, with a Hamming window size of 32ms and an overlap of 0.75\%. The signal was sampled at 16kHz. Only the STFT magnitude was considered and for the reconstruction of the time-domain signal we used the noisy phase.
Only 256 frequency bins were taken into account owing to symmetry, and ignoring the high frequency bin in order to use an exact power of two that allows a simpler network.
The data were divided into groups of 256 time bins each, so each group formed a $256 \times 256$ image.

\subsection{Results}
We compared our approach to the methods that competed in the REVERB challenge in the category ``single channel utterance based". In this category, enhancement is required to be solely dependent on the specific utterance (and the training data), without using the other utterances with same conditions (e.g. same room). Each utterance was thus separately enhanced. The results of the simulated test data for far and near microphones are described in Table~\ref{table:sim_far} and Table~\ref{table:sim_near}, respectively. It is evident that for the far microphone (where reverberation conditions are harsher), U-Net with asymmetric filters exhibits better LLR and FWSegSNR objective measures than the other methods. Even for the CD objective measure, this method was the best in room 3. The best performance improvement was achieved for the FWSegSNR objective measure, regardless of the room type. For the near microphone, the regular U-Net and asymmetric U-net outperformed the other methods in most of the rooms for the CD, LLR and FWSegSNR objective measure, whereas the differences between the regular and asymmetric U-Net were negligible. In addition, for the CD objective, the differences between the best result and U-Net were very small, for both the far and near microphones. However, in terms of SRMR our method performed less well. In spite of the failure on the SRMR in the simulated data, the results for real recordings confirmed the success of the asymmetric U-Net method as compared to the others (Table~\ref{table:real}). The GAN
had increased its FWSegSNR score, along with its CD in room 3, with almost no injure in other measurements.
A subjective evaluation campaign was not carried out. The interested reader may check a few examples in our website:\\ \texttt{\justify www.eng.biu.ac.il/gannot/speech-enhancement/speech-dereverberation-using-fully-convolutional-networks/}.

\section{Conclusion}

In this study we presented a deep learning approach for enhancing noisy and reverberant speech based on image-to-image processing of the log-spectrogram of the reverberant
speech signal. Working directly on the image representation enabled us to explicitly model typical time-frequency patterns. We obtained significant improvements on both real and simulated data compared to previously suggested approaches.

\section{Acknowledgment}

The authors would like to thank Daniel Michelsanti and Zheng-Hua Tan for sharing their GAN code.

\balance
\bibliography{citation}
\bibliographystyle{IEEEtran1}

\end{document}